# Statistical Analysis of Metrics for Software Quality Improvement

Karuna P[1], Divya MG[2], Mangala N[3]

[1]Centre for Development of Advanced Computing
Knowledge Park, Old Madras Road, Bangalore
karunap@cdac.in
[2]Centre for Development of Advanced Computing
Knowledge Park, Old Madras Road, Bangalore
divyam@cdac.in
[3]Centre for Development of Advanced Computing
Knowledge Park, Old Madras Road, Bangalore
mangala@cdac.in

## Abstract

*Software product quality can be defined as the features and characteristics of the product that meet the user needs. The quality of any software can be achieved by following a well defined software process. These software process results into various metrics like Project metrics, Product metrics and Process metrics. Software quality depends on the process which is carried out to design and develop software. Even though the process can be carried out with utmost care, still it can introduce some error and defects. Process metrics are very useful from management point of view. Process metrics can be used for improving the software development and maintenance process for defect removal and also for reducing the response time.*

*This paper describes the importance of capturing the Process metrics during the quality audit process and also attempts to categorize them based on the nature of error captured. To reduce such errors and defects found, steps for corrective actions are recommended.*

## 1. Introduction

The quality of software is of utmost importance in the field of software engineering. Software quality also depends on the process which is carried out to design and develop the software. Even after the process is followed with utmost care, the errors and defects may still exist. The quality of a software product is mainly determined by the quality of the process used to build it. Measurement and analysis will help in determining the status of the software process in terms of whether the process is followed and the functioning is as intended. Verification is the similar type of control from the management perspective. To meet such goals, quality audit for software process are conducted time to time. By measuring the errors and defects, we can take steps to improve the process.

The improvement of process will depend on metrics captured in the lifecycle of software. Software metrics can be classified into Project metrics, Product metrics and Process metrics [1]. Project metrics are those that describe the project characteristics and assist in execution planning. Product metrics capture the properties of software like mean time to failure. Process metrics are management metrics which are used for improving the software development and maintenance process for defect removal and reducing response time of the process. Process metrics are invaluable tool for an organization who wants to improve their process. Usually these process metrics are not used mostly because of uncertainty about which metrics to use, how to perform measurements and how to overcome such defects.

For software process improvement, there are many models which are available for example Capability Maturity Model (CMM) [2], Bootstrap, Personal Software Process (PSP) [3], IT Infrastructure Library (ITIL) [4], IEEE [5], Six Sigma [6] and ISO 9000 quality management system [7]. These models evaluate the software product, quality and their drawback. Moreover locally designed actions can be initiated in areas where improvement is needed. The software process must be defined and documented. In addition to the processes, standards for the different work products are defined, e.g. coding and document standards.

The rest of this paper is organized as follows. In section 2, we have presented our approach and objectives. In section 3, software process improvement models are described. In section 4, we have presented the literature review which is basis of our work. In section 5, quality practices are explained, in next section 6, categorization of errors and defects are presented. In section 7, we have presented corrective actions. In section 8, data collection methodology is explained. In section 9, results and the analysis are discussed. Future work in the same is proposed in section 9. Finally we have provided conclusion in section 10.

## 2. APPROACH AND OBJECTIVE

In this paper we have applied statistical quality assurance to the errors and defects reported during the quality audit for the year 2015 and 2016 in our organization. This has been done in view to improve the quality of software development process and hence the software products. We are presenting that, by measuring the errors and defects we can take actions to improve them. We are also presenting how each and every error and defect are grouped. There after each of them is categorized based on impact of severity like minor, moderate or serious. The data collected over a period of two years has been analyzed and presented. The analysis also describes recommended actions for the corrective action. The idea has been inspired from the software engineering practitioners Roger S Pressman and Bruce R Maxim [8].

Broadly we are trying to address 3 objectives namely quality improvement, categorizing of errors and recommendation of corrective actions.

## 3. SOFTWARE PROCESS IMPROVEMENT MODELS

Few software process models commonly followed worldwide are CMM, PSP, ITIL, IEEE and Six Sigma. Capability Maturity Model (CMM) [2], is a reference model for apprising the software process maturity into various levels [10]. The different levels of Software Engineering Institute CMM have been deliberated so that it is easy for an organization to build its quality system. CMMI aimed to advance the usability of maturity models by integrating many different models into one frame work.

Personal Software Process (PSP) [3], advocates that designers should rack the way they apply time. The quality and output of an engineer is to a great degree reliant on the process being followed. PSP is a framework that helps engineers to quantify and progress. It helps in developing personal skills and approaches by estimating, planning, and tracking performance against plans, and delivers a defined process which can be tweaked by designers [9].

IT Infrastructure Library (ITIL) [4] describes processes, procedures, tasks, and checklists which are not organization specific, but can be applied by an organization for establishing integration with the organization's strategy, delivering value, and preserving a minimum level of competency. It allows the organization to establish a baseline from which it can plan, implement, and measure. It is used to demonstrate compliance and to measure progress.

IEEE [5], standards association is a group within IEEE that develops global standards in a broad range of industries including, power, energy, biomedical, health care, information technology, robotics, telecommunication, home automation, transportation, nanotechnology, information assurance, and many more [10].

Six Sigma [6] can be used for any activity that is concerned with cost, timeliness and quality of outcomes. The ultimate objective of six sigma practice is the implementation of a measurement based strategy that focuses on process enhancement [9].

## 4. RELATED WORKS

In [11] the authors from Laboratoire de génie logiciel École Polytechnique de Montréal Montréal, Canada have presented quality evaluation methodologies for specific domains or specific techniques. Normally the software product developers select a pre-defined model, customize the features, define the metrics and estimate the quality of the software product. But in this paper the authors presents a bottom-up methodology for the quality estimate process. Also a methodology is proposed and designed for an expected quality profile. Primarily, the first step is listening to the users, and then retrieving the most important quality factors and creating a model to evaluate the expected quality of the project. The profile is formed by producing the expected users' quality expectations, and then quantifying the elicited factors by applying them to our quality evaluation model and the ISO/IEC 25000 standard.

In [12] the authors have presented the mechanism of how software engineering capabilities relate to the business performance. They have proposed a structural model including the Software Engineering Excellence indicator which consisted of deliverables, project management, quality assurance, process improvement, research and development, human resource development and customer contact.

In [13] the component based software development approach has been discussed and demonstrated. Authors have proposed quality assurance model for component based software development which includes requirement elicitation, design development, certification, customization, integration, testing and maintenance.

In [14] the author has shared how NASA's Johnson Space Center developed a 'statistical method' to determine sample size for the number of process tasks to be audited by SQA. The goal of this work is to produce a high quality product which is cost effective.

In paper [15] the enslavements between requirements and architectural components are discussed so that software defects can be mitigated.

In [16] authors have said that technological choices are fundamental for project planning, resource allocation, and quality of the final software product. For analysis they have taken open source web applications available in SourceForge. They describe tools to support project managers. The authors claim that there is need to set thumb rule to guide technological choices to increase the quality of software artefacts.

In [17], the authors have introduced the evolution of software quality model standards and the facts of SQuaRE series standards. The deficiencies of ISO/IEC 2502n software quality measurement series standards were analyzed and a road map of new reference model is proposed.

Paper [18] is related to software product quality modelling and measurement. The outcome of the research is grouped as system-level software quality models, source code element-level software quality models and applications of the proposed quality models.

Our work lays emphasis on applying statistical quality assurance to advance the quality of software products.

## 5. QUALITY PRACTICES

The International body, ISO is committed to provide requirements, guidelines, specifications so on which can be used for developing quality frameworks for products and services of small and big organizations for any kind of projects. ISO International standards ensure that products and services are reliable and of good quality. The technical committees of it comprises of relevant industry experts, consumer association, academia, NGOs and government [7].

ISO 9001:2008 standards set out the criteria for a quality management system. The standard highlights quality principles like customer focus, top management motivation and continual improvement based process approach. It can be used by any organization, large or small, regardless of its field of activity. In fact is implemented by over one million companies and organizations in over 170 countries. This standard is based on a number of quality management principles including a strong customer focus, the motivation and implication of top management, the process approach and continual improvement [7].

Our organization is ISO 9001:2008 certified. ISO 9001 process is followed for the development of software products. The ISO related activities are mainly carried out by the quality assurance team. The main role of quality assurance team is ensuring quality management system conformance, promoting customer focus, and reporting on quality management system performance. A quality manager is an employee who has been given this responsibility. Monitoring the quality objectives that have been established and reporting this to 'Management Review' is another role of the quality manager. Management review focuses more on the software process rather than the software work products.

Quality manager is also responsible for internal audit planning & management. Internal audit is the disciplined approach to evaluate and improve the effectiveness of software quality processes. The scope of internal audit is mainly risk management, control and governance of software processes. Internal audits are done by the quality assurance team to check the availability of the documents and to ensure that all the important and basic parameters were covered or not in terms of non-conformance points.

The core components of software development are Software Requirement elicitation, Design phase, Implementation phase and Testing phase. IEEE Std 1074 is Standard for Software Lifecycle which mainly covers the above listed phases. Requirements Engineering Process captures the requirements addressing the functionality, performance, attributes, constraints, human resources, hardware and software interfaces. The attributes like portability, maintainability, security are also addressed. Required standards and operating environment are listed. Process also captures boundary of the system, intended users, total users, maximum users at any one time, type of users and so on. Deliverables like system help files, manuals, documentation, source code, training and support aspects are also mentioned.

Overall the requirement process tries to bring the clarity, completeness, consistency, traceability and feasibility aspects. Any change in requirements is dealt with change management by prioritizing them. Any change is evaluated based on the feasibility and risks in achieving the new requirement. The Software Requirement Specification document is a concise document capturing the above aspects which goes through a peer review process and suggested changes are accepted or rejected after the discussion. Finally the approval of the document by the project manager becomes the baseline for the entire lifecycle of the project.

Design phase captures the design specification. It provides a high-level overview of how the functionality and responsibilities of the system were partitioned and then assigned to subsystems or components. A description of all data structures including internal, global, and temporary data structures are listed. Reference to data dictionary and data flow diagrams (DFD) are created during requirements analysis. A detailed description of each software component contained within the architecture is presented. Documents all the design attributes like performance considerations, reliability, portability, user interface, details for the preservation of products etc. Design verification is carried through the Technical Review or Design Walkthrough. Unit test cases or System test cases are prepared for the Design validation. If any additional features have been added in the Design phase, the same is reflected in the System Requirement Specification. It also captures the traceability matrix of requirements engineering. The end product of the design phase is the design document which goes through the technical peer review and approved for further implementation.

In the implementation phase, coding starts as per the assignment. Coding is carried out as per the coding guidelines. File header is included with proper name, path, version, no., description, function, and procedure names. Variable naming convention is according to standards. Inline comments are present wherever necessary, describing the current code blocks. Code is indented and readable. Functions used in more than one program units are put in the library files. The coding standard varies from the choice of programming language. If the project has adopted own standard or guidelines check are listed. The deviations from the standard/guidelines are justified.

Testing process covers the testing activities carried out at various phases of software development. Testing activities include, test planning, designing test cases, executing the test cases, evaluating the software based on test results, measuring and analyzing test data. Test cases are designed for verifying each requirement. Test cases for unit tests are identified with the input and output data. Integration testing identifies of the environment needed for integration critical modules and schedules of testing. System testing is done to validate the software product against the requirement specification. Here attributes such as external interfaces, performance, security, configuration sensitivity, co-existence, recovery and reliability are validated during this phase. A series of tests are performed to ensure that the system satisfies all its functional and non-functional requirements.

## 6. ERROR AND DEFECT CLASSIFICATION

Data collection of various software parameters and measurement provides insights to project management team and managers. The measurement is possible by first collecting quality data and then it can be compared with past data and evaluate whether improvements has occurred. The software can be measured based on Project, Product and Process and hence can be classified as Project metrics, Product metrics and Process metrics [1]. Project metrics capture defects, cost, schedule, productivity and estimation of project resources and deliverables. Product metrics measure cost, quality and time to market. Process metrics are related to quality process followed for software development. They measure the efficiency and effectiveness of various processes. Process metrics can be systematically captured from the software quality audits. Software quality audit is an independent and systematic examination for determining any deviation from the planned activities. An audit is the examination of the work products and related information to assess whether the standard process was followed or not. The data for our analysis is collected from the "Auditor Note Sheet". The collected data is analyzed based on its nature and classified into various types like erroneous specification, misinterpretation or incomplete or inaccurate documentation etc.

1. Incomplete or erroneous specifications - Any specification incompletion is captured in this category. Any deviations from the process manual or specification like approval missing, partial implementation etc are included. If any missing metrics in the specification/template is also considered as error under this category.
2. Violation of programming standards - Any deviation from standards or introduction or modification can be counted in this category.
3. Error in data representation - Any deviation from data formats as declared in specification.
4. Inconsistent competent interface - Any deviation from recommended interface related errors.
5. Error in design logic - Any deviation from committed logic eg DFDs, UML or ER diagram.
6. Incomplete or erroneous testing - Any errors and defects reported in testing by stakeholder/ customer/ third-party user etc after completion of testing.
7. Intentional deviation from specification - It relates to deviation from process manual, software requirement specification etc due to lack of suitable reasons.
8. Inaccurate or incomplete documentation - Any missing sub sections of process manual or incomplete documentation.
9. Assorted error type - Any other errors and defect not captured in above mentioned categories.

All of the above categories are further classified based on the severity of the error/defects. They are labelled as minor, moderate and serious. It is classified as minor if the error/ defect not critical to impact the process. Similarly, the defect is classified as moderate if the process is observed to be followed but cannot be evidenced. If the error or defect is observed to have major deviation from process then it is categorized as serious.

## 7. CORRECTIVE ACTION

For each of the error and defect categorized above, a corrective action is recommended as discussed below;

1. Incomplete or erroneous specifications - Effective Peer Review to be conducted.
2. Violation of programming standards - Reason to be captured for intentional violation and same to be reviewed.
3. Error in data representation - Recommend to use tools for data modelling also perform more stringent data design reviews.
4. Inconsistent competent interface - Recommend more appropriate technical reviews and trainings.
5. Error in design logic - Recommend more appropriate technical reviews and trainings.
6. Incomplete or erroneous testing - Recommend to adopt more appropriate testing methodologies with proper test plans.
7. Intentional deviation from specification - Reasons to be captured for intentional deviation and same to be reviewed.
8. Inaccurate or incomplete documentation - Recommend to use tools for documentation and reviews.

## 8. DATA COLLECTION

At C-DAC, [19] the software quality audit is conducted quarterly. Audit is conducted for every project which is in design phase, development phase and maintenance phase. Quality assurance team rolls out the schedule with date, time, project name, auditee, auditor, and venue. With this auditee will keep ready all document and details required for audit. After the audit auditor will submit "Auditor Note Sheet" to quality assurance team. Auditor note sheet contains audit errors and defects, if any. Quality assurance team publishes the entire "Auditor Note Sheet" in ISO related intranet web site where all C-DAC members have access to these Note Sheets.

Table 1- Error Categorization for year 2016

| **Error Type** | **Serious Errors** | **Moderate Errors** | **Minor Errors** |
|---|---|---|---|
| Violation of programming standards | 0 | 0 | 0 |
| Incomplete or erroneous specifications | 1 | 2 | 11 |
| Error in data representation | 0 | 0 | 0 |
| Inconsistent competent interface | 0 | 0 | 0 |
| Error in design logic | 0 | 0 | 0 |
| Incomplete or erroneous testing | 0 | 0 | 0 |
| Intentional deviation from specification | 2 | 1 | 0 |
| Inaccurate or incomplete documentation | 0 | 0 | 0 |
| Assorted error type | 0 | 0 | 0 |
| Total | 3 | 3 | 11 |

Table 2 - Error Categorization for year 2015

| **Error Type** | **Serious Errors** | **Moderate Errors** | **Minor Errors** |
|---|---|---|---|
| Violation of programming standards | 0 | 0 | 0 |
| Incomplete or erroneous specifications | 1 | 2 | 6 |
| Error in data representation | 0 | 0 | 0 |
| Inconsistent competent interface | 0 | 0 | 0 |
| Error in design logic | 0 | 0 | 0 |
| Incomplete or erroneous testing | 0 | 0 | 0 |
| Intentional deviation from specification | 3 | 1 | 0 |
| Inaccurate or incomplete documentation | 0 | 0 | 0 |
| Assorted error type | 0 | 0 | 0 |
| Total | 4 | 3 | 6 |

For our experiment we have taken 2 years data namely Year 2016 and Year 2015. Based on our quality assurance guidelines of our organization these errors and defects are grouped as serious, moderate and minor which is described in section 6. Also based on its nature every error or defect is categorized as per section 6, same is recorded in the Table 1 and Table 2. Figure1 and Figure2 capture the severity of the errors thus categorized.

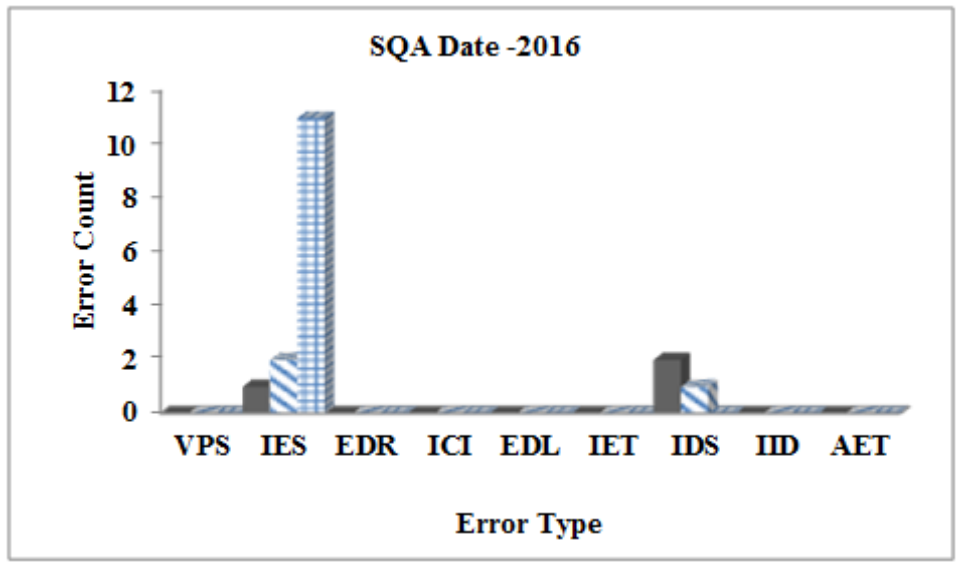

Figure 1- Severity of errors captured for year 2016

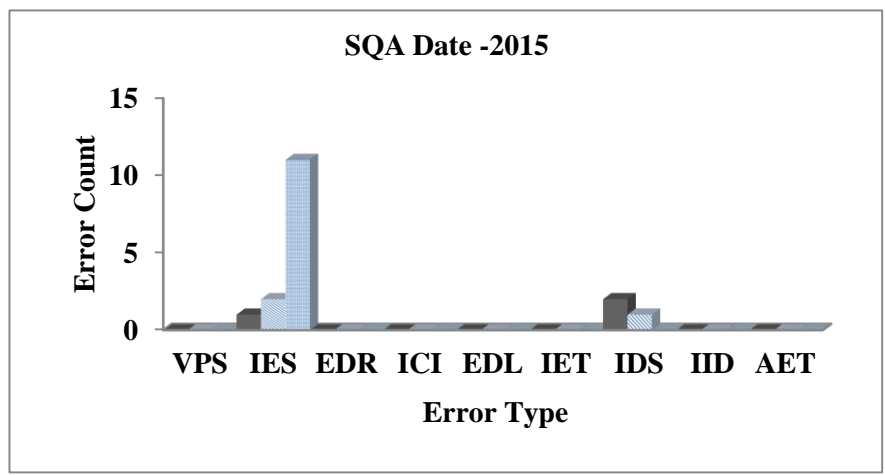

Figure 2- Severity of errors captured for year 2015

## 9. ANALYSIS AND RESULTS

Every year three internal audits and one external audit's are conducted. Internal audit is conducted by Software Quality Assurance team of C-DAC, external audit is conducted by external authorities. During the audit, auditors will recode their observation, errors and deviations. This is termed as "Non Conformity- (NC)" in "Auditor Note Sheet" statement. We have collected all the NC's reported, same is categorized as per section 6 and grouped as serious, moderate and minor. The total serious, moderate and minor errors of both the years are represented in Table 3 and Table 4. Figure3 and Figure4 projects the cumulative errors for two years.

Table 3 – Severity of Cumulative Errors

| Type of errors | Year 2015 | Year 2016 |
|---|---|---|
| Serious | 4 | 3 |
| Moderate | 3 | 3 |

| Minor | 6 | 11 |

Table 4 – Cumulative errors for 2 years

| Sl No | Year | Total errors |
|---|---|---|
| 1 | 2016 | 17 |
| 2 | 2015 | 13 |

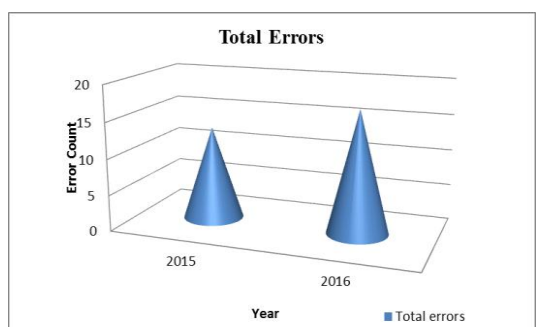
Figure 3 – Projection of errors

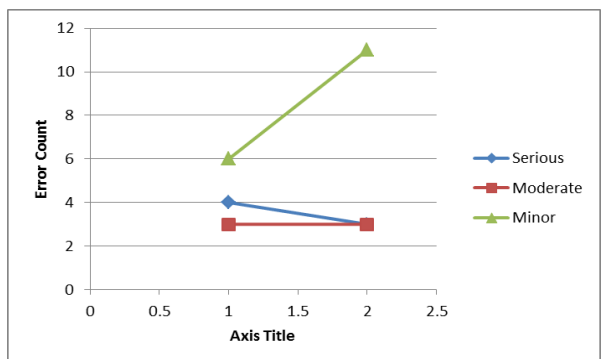
Figure 4 – Cumulative Projection of Severity errors

All the errors and defects are categorized and grouped mainly to know the statistics of software quality of projects. The data represented in Table 3 is collected from 9 projects. The projects are either in design, implementation or maintenance states. The projects belong to various domains such as distributed computing, cryptography, high performance computing, Internet of things, mobile applications etc. These projects are implemented in programming languages java, C, python and other scripting languages. Some of these are using databases.

In Table 3, it is documented that in year 2016 overall error reported was 17. Out of which 3 are serious, 3 are moderate and 11 are minor type. The one serious error was due to Incomplete or erroneous specifications- effective 'peer review process' was recommended. Remaining 2 serious errors was due to Intentional deviation from specification – reason was Work Breakdown Structure was not updated, approval was not taken in time etc. All the causes of errors were analyzed and training provided on quality process. Also, there were 2 moderate and 11 minor errors due to 'Incomplete or erroneous specifications' and one more was due to

'Intentional deviation from specification'. In both the case effective peer review process and training on quality process was recommended. Similar analysis was carried for the year 2015.

The objective of the paper is to measure the errors and defects (non conformity) of all the projects, review it and recommend the appropriate corrective action. So that the project cost will not over shoot, it can be delivered in time also the quality of the project will increase. Hence software quality of products delivered by organization improves.

## 10. FUTURE WORK

Here we are describing the work of error categorization. After collecting error and defect information, error index can be calculated. In future, we intend to calculate the phase index and error Index, which is an overall indication of improvement in software quality.

## 11. CONCLUSIONS

To improve the software quality, we collected software Process metrics. Our focus was mainly towards collecting metrics obtained through the quality control process. The errors and defects found through the software quality audits was the base data. These defects were subsequently categorized into nine types. Defects are analysed, recommendations for improving such defect are suggested.

The improvement process was suggested which mainly consist of short training on familiarity with Software Process, recommended technical reviews and group discussions for achieving the higher quality. The steps were analysed where defect occurred, identified and elaborated for stepwise corrections. Successful use case was demonstrated through an improvement program. The weak areas of defects were identified and expert help was imparted to resolve them. Remarkable improvement observed in quality of software product after implementing the recommendation to the errors and defects found.

### ACKNOWLEDGMENT

We are grateful to Ms Veena KS, Software Quality Assurance team for sharing the data and our Senior Management team for having a keen eye on improving the quality of software. We are thankful to Centre for Development of Advanced Computing (C-DAC), the premier R&D organization under the parent organization Ministry of Electronics and Information Technology (MeitY) for supporting this work.

# Authors

**Ms Karuna P** is working as Principal Technical Officer in OGGI group at C-DAC. She has received Master in Computer Applications (MCA) from Nagpur University and MS in Software Systems from BITS, Pilani. She has actively worked in the development of grid and cloud monitoring software, workflows and science gateways on Grid Computing platforms and distributed systems.

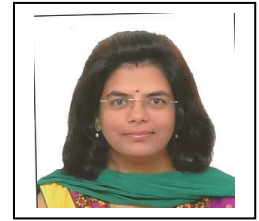

**Ms Divya MG** is working as Joint Director in OGGI group at C-DAC. Her educational qualification is B.E, M.S (SS), M.B.A. She is involved in Grid operations. Her expertize areas are telecommunication, high performance computing, gird and cloud computing. She was working as scientist in Core R&D, ITI limited.

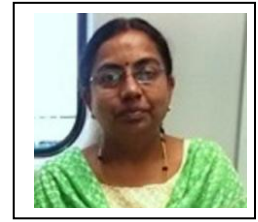

**Ms. Mangala N** is working as Associate Director, System Software Group at C-DAC. She has Master of Engineering degree in Computer Science and Engineering (CSE) and Bachelor of Engineering in CSE from Bangalore University. She has executed several projects under Heterogeneous High Performance Computing, Grid and Cloud Computing. She has actively worked in the development of Automatic Parallelizing F90 Compiler, Coral66 Compiler, Cross Assembler for special purpose microprocessor, Ada95 Compilation System, Grid Monitoring software, Grid middleware, Grid Integrated Development Environment, Workflows and Problem solving Environments, heterogeneous runtime, cluster scheduler, heterogeneous program development environment, and in parallelizing & optimizing applications on HPC and Grid Computing platforms.

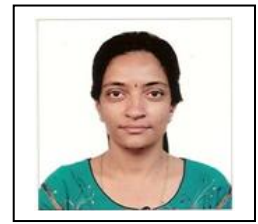